\documentclass[twocolumn]{aastex62}
\usepackage{graphicx}
\usepackage{amsmath}
\usepackage{gensymb}
\usepackage{natbib}

\graphicspath{{./}{figures/}}

\shorttitle{A Dust Trap in HD 34700}
\shortauthors{Benac et al.}

\begin{document}

\title{A Dust Trap in the Young Multiple System HD 34700}

\author{Peyton Benac}
\affil{Harvard-Smithsonian Center for Astrophysics, Cambridge, MA 02139, USA}

\author{Luca Matr\`{a}}
\affil{School of Physics, National University of Ireland Galway, University Road, Galway, Ireland}
\affil{Harvard-Smithsonian Center for Astrophysics, Cambridge, MA 02139, USA}

\author{David J. Wilner}
\affil{Harvard-Smithsonian Center for Astrophysics, Cambridge, MA 02139, USA}

\author{Mar\`{i}a J. Jim\`{e}nez-Donaire}
\affil{Observatorio Astron\'omico Nacional, Alfonso XII 3, 28014, Madrid, Spain}

\author{J. D. Monnier}
\affil{Astronomy Department, University of Michigan, Ann Arbor, MI 48109, USA}

\author{Tim J. Harries}
\affil{Department of Physics and Astronomy, University of Exeter, Exeter, EX4 4QL, UK}

\author{Anna Laws}
\affil{Department of Physics and Astronomy, University of Exeter, Exeter, EX4 4QL, UK}

\author{Evan A. Rich}
\affil{Astronomy Department, University of Michigan, Ann Arbor, MI 48109, USA}

\author{Qizhou Zhang}
\affil{Harvard-Smithsonian Center for Astrophysics, Cambridge, MA 02139, USA} 

\begin{abstract}

Millimeter observations of disks around young stars reveal substructures indicative of gas pressure traps that may aid grain growth and planet formation. We present Submillimeter Array observations of HD 34700- two Herbig Ae stars in a close binary system (Aa/Ab, $\sim$0.25 AU), surrounded by a disk presenting a large cavity and spiral arms seen in scattered light, and two distant, lower mass companions. These observations include 1.3 mm continuum emission and the $^{12}$CO 2-1 line at $\sim0\farcs5$ (178 AU) resolution. They resolve a prominent azimuthal asymmetry in the continuum, and Keplerian rotation of a circumbinary disk in the $^{12}$CO line. The asymmetry is located at a radius of $155^{+11}_{-7}$ AU, consistent with the edge of the scattered light cavity, being resolved in both radius ($72 ^{+14}_{-15}$ AU) and azimuth (FWHM = $64 ^{\circ +8}_{-7}$). The strong asymmetry in millimeter continuum emission could be evidence for a dust trap, together with the more symmetric morphology of $^{12}$CO emission and small grains. We hypothesize  an unseen circumbinary companion, responsible for the cavity in scattered light and creating a vortex at the cavity edge that manifests in dust trapping. The disk mass has limitations imposed by the detection of $^{12}$CO and non-detection of $^{13}$CO. We discuss its consequences for the potential past gravitational instability of this system, likely accounting for the rapid formation of a circumbinary companion. We also report the discovery of resolved continuum emission associated with HD 34700B (projected separation $\sim1850$AU), which we explain through a circumstellar disk.

\end{abstract}

\section{Introduction} \label{sec:intro}
In recent years, protoplanetary disks have been shown to contain substructures in the form of spiral arms, rings, cavities, and asymmetries \citep[e.g.][]{andrews18a}. In many systems, prominent spiral arms are observed in the infrared, while a more concentrated and strongly asymmetric congregation of dust grains is seen in (sub-)mm wavelengths, e.g. IRS 48, V 1247 Ori, LkHalpha330, MWC 758, HD142527, and HD 135344B \citep{vandermarel, follette15, kraus17,fuente17, akiyama16, isella13, marino15, casassus12, casassus13, casassus15, cazzoletti}.  These (sub-)mm azimuthal asymmetries  are considered evidence of  dust trapping, as proposed by \citet{whipple72}, \citet{1977}, and \citet{bargesommeria}. When a local pressure maximum arises, large dust grains drift toward it and grow. By preventing radial drift of large grains into the star, these pressure maxima become a site of efficient dust growth, potentially leading to planetesimal formation.  Evidence of strong azimuthal dust trapping in a protoplanetary disk was first observed by \citet{vandermarel} in (sub-)mm observations of the system Oph IRS 48. Because thermal emission observations typically probe grains of a size comparable to the observing wavelength, (sub-)mm observatories like the Submillimeter Array (SMA) are uniquely able to image the compact emission that arises from the enhanced clumping of large dust grains \citep{beckwith90, testi03}.\\
HD 34700 is a multiple system comprised of a binary Herbig Ae system (HD 34700AaAb, with a semimajor axis of $a \sim 0.25$ AU based on the spectroscopic observations of \citet{torres}) and a distant tertiary companion HD 34700B, located at a projected distance of $\sim$1850 AU from the central binary, as well as another companion HD 34700C at a projected separation of 3300 AU \citep{sterzik05}. The central binary is composed of two close, young stars, of age $\sim 5$ Myr and masses $\sim 2 M_\odot$, orbiting with a period of approximately 23 days and an eccentricity of $\sim 0.25$ \citep{torres, monnier}. The new distance determination from \citet{gaia} revised the previously uncertain distance to the system to be $356.5\pm6.1$ pc, which significantly reduced the estimate of the age of the system \citep{monnier}. This revised age indicates that the disk is a transition disk around a pair of young stars, not a debris disk around an older main-sequence star. The shape of the spectral energy distribution suggested a cavity around the central binary, which was confirmed by NIR GPI observations presented in \citet{monnier}. These near-IR scattered light observations resolved the disk around the central binary for the first time, unveiling a large, $\sim 175$ AU cavity and multiple spiral arms in a circumbinary disk with an inclination of $\sim 42 ^\circ$ and elongation along a position angle of $\sim 69 ^\circ$ E of N \citep{monnier}. New Subaru/SCExAO+CHARIS observations \citep{Uyama20} yielded $JHK$-band images that confirmed the ring structure and spirals imaged with GPI in \citet{monnier}. These observations revealed darkening on the ring and spirals, and were used to set limits on the mass of potential companions \citep{Uyama20}. These scattered light observations trace the (sub-)micron grain population in the upper layers of the disk \citep[e.g.][]{juhaszrosotti18}. These small grains are least susceptible to congegrating in pressure maxima \citep{1977}. Therefore, we use (sub-)mm observations to probe the distribution of larger, mm-sized grains that lie closer to the circumbinary disk's midplane.\\
In this paper, we present new SMA subarcsecond resolution observations of the HD 34700 system. In Section \ref{sec:obs}, we describe the observations and the data products. In Sections \ref{sec:results} and \ref{sec:model}, we present results of the imaging and model fits to the visibilities that characterize the observed features.  In Section \ref{sec:disc}, we discuss interpretations and possible origins of these features. In Section \ref{sec: conc}, we summarize the conclusions.\\

\section{Observations} \label{sec:obs}
\begin{deluxetable*}{cccc}
\tablecaption{SMA Observational Parameters \label{tab:obstable}}
\tablewidth{0pt}
\tablehead{
\colhead{Parameter} &  \colhead{2019 Jan 08} & \colhead{2019 Mar 02} & \colhead{2019 Mar 04}}
\decimalcolnumbers
\startdata
No. Antennas & 8 & 7 & 8\\
Configuration & SUB & VEX & VEX \\
$\tau_{225}$ & 0.04 & 0.04 & 0.03 \\
Min/Max Baselines & 9 to 69 meters & 68 to 508 meters & 68 to 508 meters \\
Gain Calibrators & 0510+180, 0532+075 & 0509+056, 0532+075 & 0509+056, 0532+075 \\
Passband Calibrators & 3c84 & 3c279, 3c84 & 3c279, 3c84 \\
Flux Calibrator & Uranus & Uranus & Uranus \\
Synthesized Beam FWHM & 4\farcs4 $\times$ 3\farcs4 & 0\farcs58 $\times$ 0\farcs41 & 0\farcs58 $\times$ 0\farcs41 \\
\enddata
\end{deluxetable*}

We used the Submillimeter Array (SMA)\footnote{The Submillimeter Array is a joint project between the Smithsonian Astrophysical Observatory and the Academia Sinica Institute of Astronomy and Astrophysics, and is funded by the Smithsonian Institution and the Academia Sinica.} \citep{homoranlo2004} to image the continuum, $^{12}$CO and $^{13}$CO J=2-1 emission in the HD 34700 system at high angular resolution. Observations in the very extended (VEX) configuration were performed on March 2 and March 4, 2019. Subcompact (SUB) observations, which did not cover the $^{13}$CO line, were performed on January 7, 2019 to provide additional information about the possible presence of large scale structures filtered out by the long VEX baselines. A summary of the observational parameters is available in Table \ref{tab:obstable}. Observations are centered at R.A. 05:19:41.4097, Dec +05:38:42.8037 (J2000), which is the expected position of the central binary based on Gaia coordinates and proper motion extrapolated to the date of observation \citep{gaia}. The spectral setup of the VEX observations was centered at the local oscillator frequency of 225 GHz. The frequencies covered in the spectral observations were 229 to 237 GHz in the upper sideband and 213 to 221 GHz in the lower sideband. The main spectral line of interest in these observations was $^{12}$CO J=2-1, located at a rest frequency of 230.538GHz. The primary beam FWHM in these observations was $55\farcs$. \\ 

Calibration of the interferometric visibilities was performed using the MIR\footnote{\url{https://www.cfa.harvard.edu/~cqi/mircook.html}} software package. Corrections were made for system temperature, and the calibrators listed in Table \ref{tab:obstable} were used for gain, passband, and flux calibration.
Imaging and deconvolution were performed using the \textit{tclean} task in the Common Astronomy Software Applications package \citep[CASA;][]{mcmullin07}.
Before imaging the continuum, the visibilities at the frequency of the known CO transition were flagged and excluded. The continuum imaging was performed using a natural weighting scheme with no taper. The continuum image has an rms of 0.19 mJy~beam$^{-1}$ with a synthesized beam size of $0\farcs58 \times 0\farcs41$. At the Gaia-determined distance of 356.5 pc, this translates to a resolved physical scale of 210 $\times$ 150 AU. When referring to a radius, we always refer to radii in the orbital plane (i.e. deprojected), rather than on-sky radii. \\
In order to image the CO emission, the continuum emission was subtracted using the \textit{uvcontsub} task in CASA. The $^{12}$CO and $^{13}$CO lines were also cleaned with natural weighting with no taper. The $^{12}$CO ($^{13}$CO) cube was produced with a channel width of 0.73 (1) km~s$^{-1}$ and a single channel RMS noise level of 30 (25) mJy~beam$^{-1}$ for a synthesized beam of $0\farcs56\times0\farcs40$ ($0\farcs57\times0\farcs40$).
  
  The images presented use the VEX observations only, but the SUB data are included in the visibility modeling performed in Section \ref{sec:model}. As shown by our visibility modeling in Section \ref{sec:model}, these VEX observations recover all of the flux detected in the very compact configuration SUB observations. Therefore, these images provide the highest angular resolution for the resolved morphologies, without the complexities introduced into the synthesized beam shape by the large gap in $u,v$ coverage between the VEX and SUB baselines. 
\begin{figure}[h]
\label{fig:dust}
\centering
\includegraphics[scale=0.39]{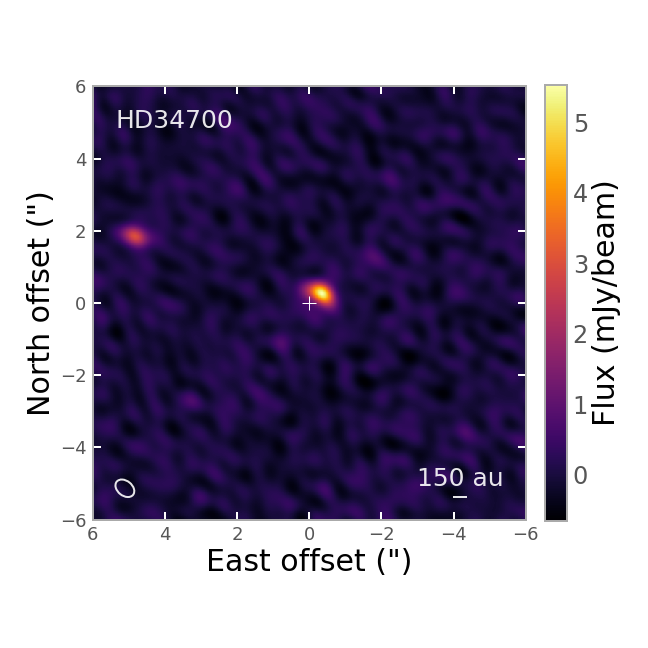}
 \vspace{-15mm}
 \caption{Cleaned dust continuum image ($\lambda=1.3$mm) of the HD 34700 multiple system obtained with natural weighting using the VEX data. The white ellipse in the lower left represents the synthesized beam size of $0\farcs58 \times 0\farcs41$. The plus symbol denotes the location of the central binary as determined from Gaia coordinates and proper motion, corresponding to the phase center of our observations.}
\end{figure}

\begin{figure}[h]
\centering
\includegraphics[scale=0.37]{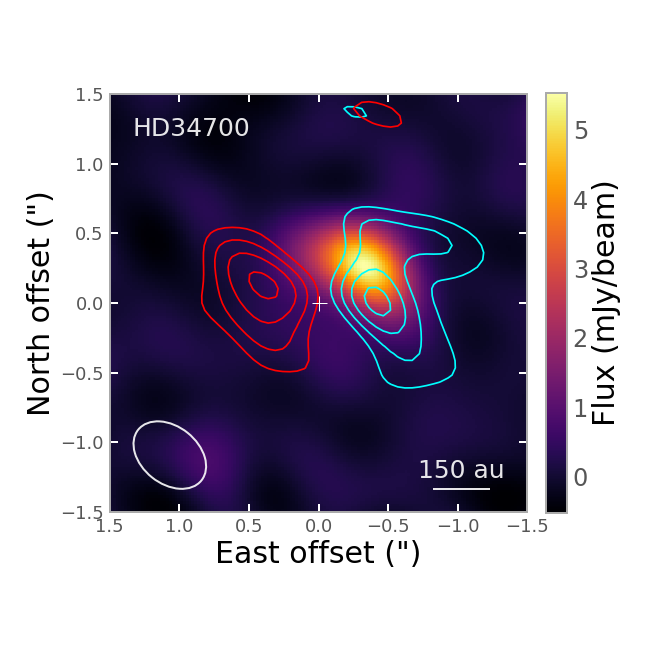}

\vspace{-15mm}
\caption{Contours represent the CO J=2-1 moment-0 emission from blue- and red- shifted velocity with respect to the HD 34700 AaAb binary's systemic velocity of 21 km~s$^{-1}$ in the heliocentric reference frame \citep{pourbaix04}. Blueshifted emission includes emission from 17 to 21 km~s$^{-1}$, and redshifted emission from 21 to 25 km~s$^{-1}$. Contours are drawn at 4, 6, 8, and 10 times the RMS of the CO moment-0 map (49 mJy~beam$^{-1}$ km~s$^{-1}$), and are overlaid on a zoomed-in version of the dust continuum image of Figure \ref{fig:dust}. The continuum beam size of $0\farcs58 \times 0\farcs41$ is shown in the ellipse at lower left.}
\label{fig:contours}
\end{figure}
\section{Results} \label{sec:results}

\begin{figure}
     \centering
     \includegraphics[scale=0.4]{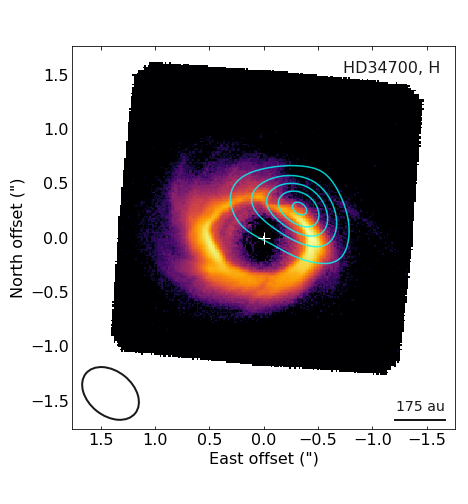}
     \caption{H band GPI image of the circumbinary disk around HD 34700AaAb as presented in \citet{monnier}, overlayed with contours of the SMA 1.3mm continuum observations. The GPI image shows the radial Stokes component Q$_\phi$ \citep[as defined in e.g.][]{schmid06, monnier} plotted on a logarithmic color scale. Contours represent 3, 7, 11, 15, and 19 times the continuum RMS of 0.19 mJy~beam$^{-1}$. The plus denotes the expected location of the central binary. The continuum beam size of $0\farcs58 \times 0\farcs41$ is shown in the ellipse at lower left.}
     \label{fig:overlay_NIR_SMA}
 \end{figure}

Figure \ref{fig:dust} shows the SMA 1.3mm continuum image of the HD 34700 system. This shows prominent emission next to, but offset from, the central binary which was located at the phase center of our observations, corresponding to the [0,0] location. The continuum emission is detected with a peak signal-to-noise ratio (SNR) of 30.6 and a total integrated flux of $7.4 \pm 1.6$ mJy, including a 20\% flux calibration uncertainty added in quadrature. Given the SNR of the azimuthally asymmetric emission, the contrast against undetected emission (if any) at other azimuthal locations is at least 10.2. The dust continuum emission around the central binary overlaps with the spiral structures observed in scattered light by \citet{monnier}, as shown in Figure \ref{fig:overlay_NIR_SMA}.\\
We also detect significant emission at a location consistent with that of the distant companion HD 34700B, as shown in Figure \ref{fig:dust}. This emission has a peak SNR of 16.7, and an integrated flux of $3.7 \pm 0.9$ mJy, including the same 20\% flux calibration uncertainty. This emission is likely to originate from a previously unknown protoplanetary disk around the tertiary companion HD 34700B.\\
The contours in Figure \ref{fig:contours} show the integrated intensity $^{12}$CO emission from blue- and red-shifted velocities, with respect to the star's velocity of 21 km~s$^{-1}$ in the heliocentric reference frame \citep{pourbaix04}. This emission is consistent with the expected velocity pattern of a Keplerian disk, and is well-centered on the location of the central binary as determined from optical observations \citep{gaia}. This confirms the proper positioning of the central binary at phase center, and therefore confirms that the offset of the continuum emission shown in Figure \ref{fig:dust} must have a physical origin and is not due to an astrometric error. The $^{12}$CO emission has a total integrated flux of $7.2 \pm 1.4$ Jy km~s$^{-1}$. 
No $^{12}$CO emission is detected at the location of HD 34700B, and no $^{13}$CO emission is detected around both HD34700AaAb and B. To set an upper limit on the $^{13}$CO integrated line flux around AaAb, we spatially integrate the cube over the same region where $^{12}$CO is detected (assuming they would be co-located), to obtain a 1D spectrum with an RMS noise level of 93 mJy in a 1 km~s$^{-1}$ channel. We then calculate a 3$\sigma$ spectrally integrated $^{13}$CO flux upper limit of 0.79 Jy km~s$^{-1}$ by assuming that the two isotopologues have the same linewidth. \\
To estimate the spatial extent of the CO emission, we measure the spatial centroid of the highest velocity channels where $^{12}$CO is detected using the \textit{imfit} CASA task. Assuming a position angle for the disk major axis of $69^\circ$ \citep{monnier}, the CO is detected down to a separation of $102 \pm 10$ AU from the central binary, which is significantly smaller than the radial location of the trap and of the bright IR ring as imaged in \citet{monnier}. This is consistent with other transition disks with large cavities, where CO gas is detected interior to the cavity seen in dust emission \citep[e.g.][]{Bruderer14, vdm18}, as predicted by models of massive planet-induced gaps \citep[e.g.][]{pinilla2012}. Higher resolution line observations are needed to study the detailed interplay of gas and dust dynamics at the cavity edge. \\

\subsection{Modeling} \label{sec:model}

 \begin{figure*}
     \hspace{-25mm}
     \includegraphics[scale=0.38]{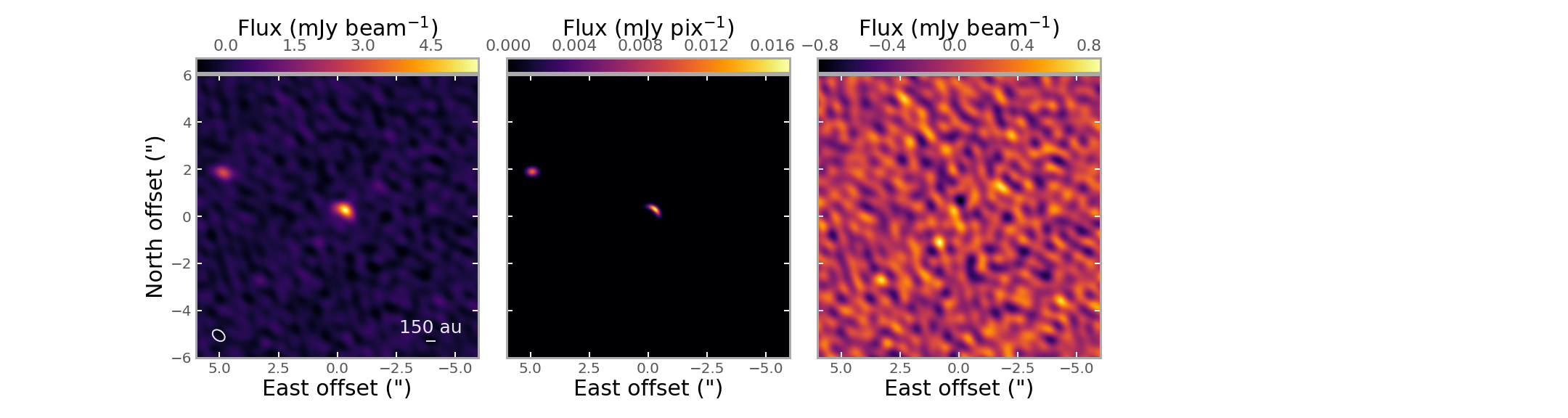}
     \vspace{-6mm}
     \caption{\textit{Left:} Natural-weighted image of VEX observations of HD 34700. \textit{Center:} Best-fit model as determined by MCMC fitting, using parameter values shown in Table \ref{tab:model}. \textit{Right:} Residual of the best-fit model.}
     \label{fig:3panel}
 \end{figure*}

The asymmetric emission detected near the central AaAb binary was modeled as a geometrically thin radially Gaussian ring with an azimuthal asymmetry, according to equation \ref{eq:gauss}, following the vortex prescription of \citet{lyralin13} and successfully used in the modeling of higher resolution ALMA observations of the dust trap around HD 135344B \citep{cazzoletti}.
\begin{equation}
    \label{eq:gauss}
    I(r,\theta) = A e^{\frac{-(r-r_c)^2}{2\sigma_r^2}} e^{\frac{-(\theta-\theta_c)^2}{2\sigma_\theta^2}}
\end{equation}
This model only includes the asymmetric crescent-shaped signal seen in the center panel of Fig. \ref{fig:3panel}, not a full circumbinary disk.
The dust emission around the companion HD 34700B was modeled as a Gaussian centered on the star with an offset from the phase center of our visibility data. We create model images from these intensity distributions. We used the \textit{GALARIO} software package \citep{galario} to Fourier transform the model image and sample the resulting visibility function at the same u-v coordinates that were sampled in both our SUB and VEX SMA observations. The SUB and VEX data were fitted simultaneously using a model that included both the AaAb and B emission regions.

Forward modeling was performed on this data using Markov Chain Monte Carlo (MCMC) as implemented in  the \textit{emcee} package \citep{emcee}. The value of $\chi^2$ calculated by comparing the sampled data and the model visibilities was used to calculate the likelihood function, assumed to be $ \propto e^{{-\chi^2}/2}$. Uniform priors were chosen to enable the MCMC walkers to explore a wide but plausible range of the parameter space, with the uniform prior ranges being, in general, much broader than the width of the converged posterior probability distributions. The exception is the geometric parameters, for which the parameter space is bound by definition (e.g. inclination is defined between $0^\circ$ and $90^\circ$, as is position angle between $-180^\circ$ and $180^\circ$.) Fourteen parameters, listed in Table \ref{tab:model}, were fit using \textit{emcee}. Beyond the geometric parameters of the Gaussian models and the flux, the offset of the SUB observations from the VEX observations in RA and Dec were also fitted as nuisance parameters. 
The position angle of the disk around the inner AaAb binary is fixed to 69$^\circ$, as reported in \citet{monnier}. \\
 Table \ref{tab:model} contains the parameter values that yielded the best fit to the data, listed as the $50 \pm 34$th percentile of the posterior probability distributions. The reduced $\chi^2$ of this model is 1.02. We estimate the dust emission to be at a distance of $\sim 155$ AU from the central binary and to lie on a plane that is inclined by less than $47 ^\circ$ from face on. This places the emission at a location consistent with the cavity described by \citet{monnier} based on near-IR imaging, as shown in Figure \ref{fig:overlay_NIR_SMA}, and at an inclination consistent with the value of $42 ^\circ$ inferred from the same NIR observations.
 
\begin{figure*} 
    \centering
    \vspace{-20mm}
    \includegraphics[scale=0.62]{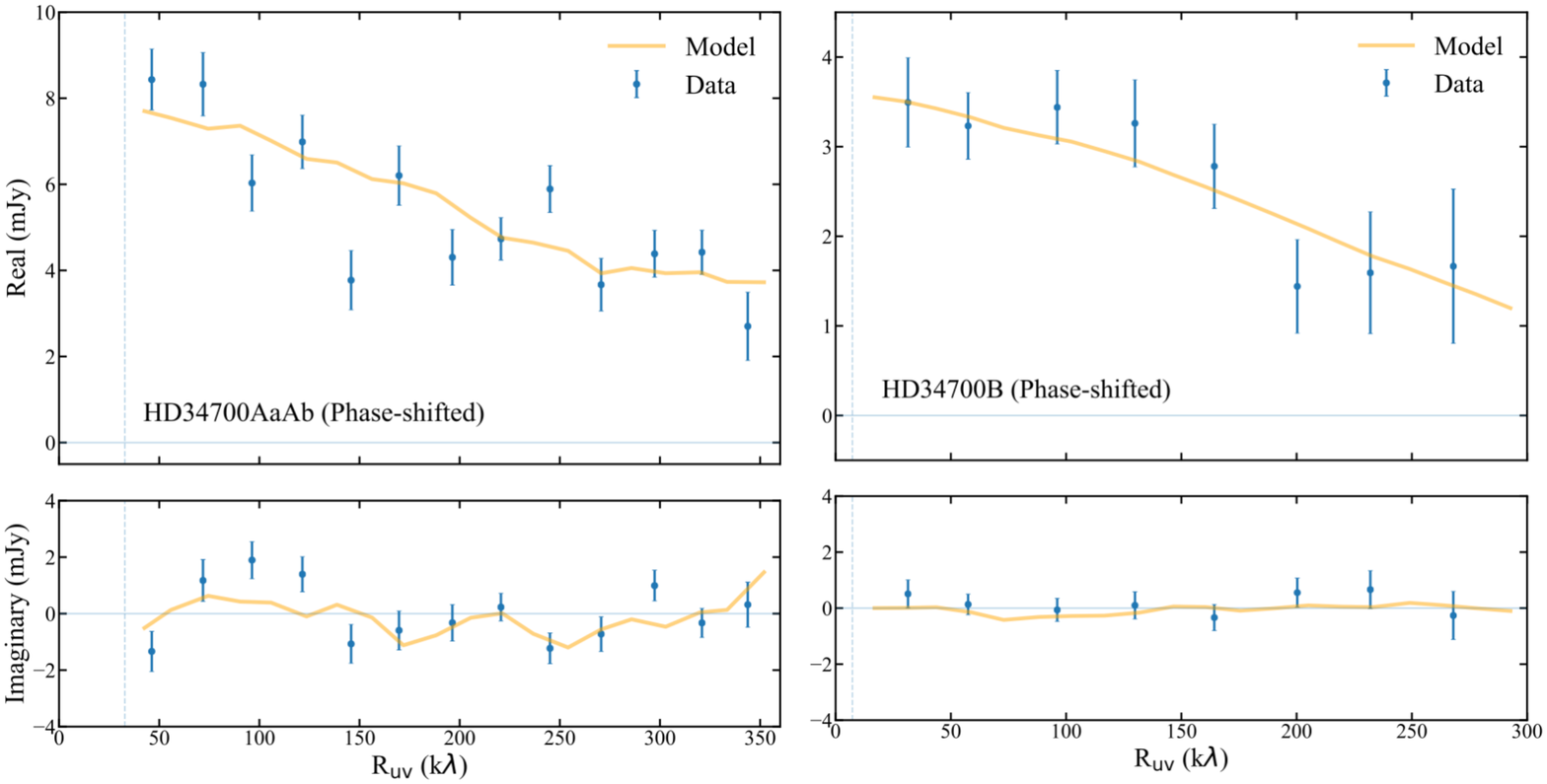}
    \vspace{-24mm}
    \caption{Real (top) and imaginary (bottom) part of the observed visibilities as a function of \textit{u-v} distance. \textit{Left:} Visibility function of the emission around the inner binary HD 34700AaAb, after subtracting the best-fit emission around B, and phase shifting to the on-sky location of the offset emission region. \textit{Right:} Deprojected visibility profile of the emission around the companion HD 34700B, after subtracting the best-fit emission around AaAb, and phase shifting to the on-sky location of HD 34700B. The best-fit inclination and PA (Table \ref{tab:model}) of B were used to apply the deprojection, following the method of e.g. \citet{Hughes2007}. For both HD 34700AaAb and B, the real part is decreasing significantly as a function of u-v distance, indicating that the emission is spatially resolved by the SMA baselines.}
    \label{fig:uvprofs}
\end{figure*}

Our visibility fitting indicates that the dust emission around both HD 34700AaAb and the companion HD 34700B are spatially resolved, with the trap extending 72 AU in radius and $64^\circ$  in azimuth, and the HD 34700B emission having a Gaussian FWHM diameter of $\sim 102$ AU. This is confirmed by the observed and model visibility profiles of the emission around both HD 34700AaAb and B (Fig. \ref{fig:uvprofs}), showing a significant decrease in their real part as a function of \textit{u-v} distance. Our fitting also confirms that the centroid of the companion emission is consistent with the known location of HD 34700B from previous optical observations \citep[e.g.][]{sterzik05}. \\
Additionally, the close matching between the modeled value for the total integrated flux of the central emission and the observed value (modeled = $7.9 \pm 0.2$ mJy, observed = $7.4 \pm 1.6$ mJy) indicates that the VEX observations did not miss significant large-scale structure due to the lack of short baselines.

\begin{deluxetable}{cc}
\tablecaption{Best-Fit Parameter Values Summary \label{tab:model}}
\tablewidth{0pt}
\tablehead{\colhead{Parameter} & \colhead{Value}}
\startdata
Trap Radius ($r_c$) [AU] & $155^{+11}_{-7}$\\
Trap Radial FWHM [AU] & $72^{+14}_{-15}$\\
Azimuthal Peak Location* ($\theta_c$) & $-110^\circ \pm 2$ \\
Azimuthal FWHM & $64 ^{\circ +8}_{-7}$\\
Total Flux [mJy] & $7.9 \pm 0.2$\\
Inclination of Trap$^\dagger$ & $< 47 ^\circ$\\
Companion Flux [mJy] & $3.6 \pm 0.2$\\
RA Offset of Companion [arcsec] & $4.92 \pm 0.02$\\
Dec Offset of Companion [arcsec] & $1.87 \pm 0.01$\\
Companion FWHM [AU] & $136^{+21}_{-22}$\\
Position Angle of Companion & $89^{\circ +9}_{-8}$\\
Inclination of Companion$^\dagger$ & $> 15 ^\circ$\\
RA Subcompact Offset [arcsec] & $0.09\pm 0.03$\\
Dec Subcompact Offset [arcsec] & $0.05\pm 0.04$\\
\enddata
\tablecomments{Offsets and distances are measured from the estimated position of the central binary, corresponding to the phase center of the VEX observations. Reported uncertainties correspond to the 16th and 84th percentiles of the posterior probability distribution.\\
* - measured where the positive angular direction is counter-clockwise from the positive x-axis in the orbital plane.\\
$\dagger$ - Posterior probability distributions for both the inclination of the central disk and the companion disk were non-Gaussian in shape and skewed toward 0$^\circ$ and 90$^\circ$ respectively and therefore are reported as $3 \sigma$ upper and lower limits.}
\end{deluxetable}

\subsection{Dust Mass Estimate}
The dust emission around the central binary is resolved, as supported by the modeling in Section \ref{sec:model} and proven by the $u-v$ profiles shown in Fig. \ref{fig:uvprofs}. To calculate the dust mass around the central binary and around HD 34700B, we assume that the dust grains act as modified blackbodies and emit according to the Planck function. We can then derive a total dust mass from the observed continuum emission assuming that the dust is optically thin, 
\begin{equation} \label{eq:dustmass}
M_{\rm dust} = \frac{F_\nu d^2}{\kappa_\nu B_\nu T_{dust}}
\end{equation}
where $F_\nu$ is the measured flux and $d$ is the distance to the system. $\kappa_\nu$ is the dust grain opacity and is assumed to be $10$ cm$^2$g$^{-1}$ at 1000 GHz and scaled to the frequency of our observations (230 GHz) using an opacity power-law index $\beta = 1$ \citep{cazzoletti, beckwith90}.\\
For temperatures between 10 and 75 K, we obtain optically thin dust masses between $5.7 \pm 1.2$ and $72 \pm 15 M_\earth$ for the emission around the central binary, and $2.8 \pm 0.7$ and $35 \pm 9 M_\earth$ for the emission around HD 34700B.\\
To check our optically thin assumption, in the absence of further substructure within the beam of our observations, we can estimate the optical depth from the observed peak intensity. For dust, the optical depth $\tau_\nu$ is related to the intensity $I_\nu$ and the predicted Planck distribution $B_\nu$ by
\begin{equation} \label{eq:depthdust}
    I_\nu = (1-e^{-\tau_\nu}) B_\nu
\end{equation}
For our observed peak continuum intensity around the central binary $I_\nu$ of 5.5 mJy~beam$^{-1}$ and a temperature of 30 K, we find that $\tau_\nu \approx 0.02$, which would indicate that the dust emission around the central binary is optically thin. Additionally, the peak flux of the emission around HD 34700B is 2.95 mJy~beam$^{-1}$, which, for an assumed temperature of 30 K, yields an optical depth of 0.1. Therefore all dust emission in this system is most likely to be optically thin as long as no further substructure is present within the observed emission.

\subsection{CO Gas Mass Estimate} \label{ss:comass}
In the optically thin approximation, we can also estimate a CO gas mass $M_{\rm CO}$ from our measured CO integrated flux $F_{2-1}$
\begin{equation} \label{eq:COmass}
    M_{\rm CO} = \frac{4 \pi m d^2}{h \nu_{2-1} A_{2-1}} \frac{F_{2-1}}{x_2},
\end{equation}
where $m$ is the mass of the CO molecule, $d$ is the distance to the system, $\nu_{2-1}$ is the rest frequency of the line, $h$ is Planck's constant, $A_{2-1}$ is the Einstein A coefficient for the transition, and $x_2$ is the fraction of CO molecules that are in the upper energy level of the transition. The $^{12}$CO mass calculation is performed assuming a temperature of 38 Kelvin, as derived at the end of this Section. The fractional population, $x_j$, depends on the relative excitation of energy levels within the $^{12}$CO molecule. Assuming LTE, levels are populated according to the Boltzmann distribution, following equation \ref{eq:frac}, using the Einstein A coefficient value of $A_{2-1} = 6.91 \times 10^{-7}$ sec$^{-1}$ \citep{lamda} which yields a fractional population of the second energy level to be $x_2 = 0.309$, calculated from \begin{equation}\label{eq:frac}
    x_j = \frac{N_j}{N_{tot}} = \frac{g_j}{Z}e^{-E_j/kT_k}
\end{equation}
where $g_j$ is the degeneracy of the j\textsuperscript{th} level, $E_j$ is the energy of that level, and $Z$ is the partition function defined as $Z = \sum g_i e^{-E_i/kT_k}$. The optically thin calculation leads to a $^{12}$CO gas mass of $0.020 \pm 0.004 M_\earth$. 
This should be considered a lower limit if the emission is instead optically thick. \\
To test the latter, we here attempt to constrain the optical depth of the $^{12}$CO emission. This can be estimated using the definition of optical depth:
\begin{equation} \label{eq:CO_tau}
    \tau_{\rm CO} = \frac{h \nu}{4 \pi \Delta \nu} (x_1 B_{12} - x_2 B_{21}) N_{\rm CO}
\end{equation}
where $\Delta \nu$ represents the line width due to Doppler broadening, $B_{12}$ and $B_{21}$ are the Einstein B coefficients, $N_{\rm CO}$ is the column density of CO molecules, and $x_1$ and $x_2$ are the fraction of molecules in the 1st and 2nd rotationally excited energy levels as calculated from equation \ref{eq:frac}. We calculate the column density by approximating the region of CO emission to be a uniform density, axisymmetric disk with the radius and radial width of the trap ($\sim 155$ AU with a width of 72 AU, as determined in section \ref{sec:model}) at the same inclination as the trap (24$^\circ$, consistent with the constraint of $<47^\circ$ found in \ref{sec:model}). Using the measured CO mass of $\sim 0.02 M_\earth$, the optical depth is approximately 19, which would indicate that the CO gas is optically thick and therefore that our mass (and optical depth) is an underestimate. \\
The $3\sigma$ integrated line flux upper limit for $^{13}$CO is 0.79 Jy km s$^{-1}$. This is 9.1 times lower than our $^{12}$CO integrated line flux of $7.2\pm1.4$ Jy km s$^{-1}$. Using this lower limit on the $^{12}$CO/$^{13}$CO line ratio, we can use the following expression to crudely estimate the optical depths (as used in \citet{lyo11}).
\begin{equation}
    R = \frac{T_R(^{12}CO)}{T_R(^{13}CO)} = \frac{1-e^{-\tau_{12CO}}}{1-e^{-\tau_{13CO}}} = \frac{1-e^{-\tau_{12CO}}}{1-e^{-\tau_{12CO} / X}}
\end{equation}
This yields a maximum optical depth of 7 for $^{12}$CO and of 0.1 for $^{13}$CO, assuming that the isotopes share the same spectra-spatial morphology and excitation conditions, and that the interstellar value of $X=\frac{^{12}CO}{^{13}CO}=60$ \citep{wilsonrood94} applies.\\
Because the $^{12}$CO gas is optically thick, we can assume $I_\nu \approx B_\nu (T)$ and use the peak intensity to calculate a temperature estimate for the $^{12}$CO gas. The peak single-channel intensity was 0.24 Jy/beam; using this, we calculate the $^{12}$CO temperature at the peak intensity location to be 38 K.
\\
Assuming that the $^{13}$CO emission is optically thin, we can calculate an upper limit on its mass using Equation \ref{eq:COmass}. This yields a value of $3.5 \times 10^{-3} M_\earth$ of $^{13}$CO at a temperature of 38 Kelvin. By combining this with the interstellar ratio of $X=\frac{^{12}CO}{^{13}CO}=60$ \citep{wilsonrood94}, we can also calculate an upper limit on the $^{12}$CO gas mass. This yields a value of $0.2 M_\earth$ of $^{12}$CO. We can therefore conclude that the total CO gas mass is between the lower limit of $0.02 M_\earth$ (determined from only $^{12}$CO observations) and the upper limit of $0.2 M_\earth$ (determined from both $^{12}$CO and $^{13}$CO observations.)
\\

\section{Discussion} \label{sec:disc}
The HD 34700 system consists of a tight, eccentric ($e \approx 0.25$) central binary of two $\sim 2 M_\odot$ stars (AaAb), and two distant companions, thought to be a part of the system due to their shared radial velocity: HD 34700B at a projected separation of 1850 AU, and HD 34700C at a projected separation of 3300 AU \citep{sterzik05, torres, monnier}. These new 1.3 mm SMA observations show a Keplerian disk of CO gas and an azimuthally asymmetric gathering of dust grains $\approx 1$mm in size around the central binary, as well as dust emission presumed to be a disk consistent with the location of HD 34700B.\\ 
Previous observations of this system by \citet{monnier} at NIR wavelengths showed a disk with spiral arms and a broad cavity around the central AaAb binary. These observations place the HD 34700 system among a growing list of systems shown to have a strongly asymmetric distribution of large dust grains as imaged at (sub)mm wavelengths  \citep[e.g. Oph IRS 48, V 1247 Ori, LkHalpha330, MWC 758, and HD 135344B][]{vandermarel, follette15, kraus17,fuente17, akiyama16, isella13, marino15, cazzoletti}, some of which also show a large cavity and spiral features in NIR. Many causes are proposed for these continuum asymmetries and/or spiral arms, including planets, vortices, gravitational instability, and binary dynamics, which we will discuss in Section \ref{ss:causes}.\\
The azimuthally asymmetric dust emission observed around the AaAb binary is evidence for dust trapping taking place in this circumstellar disk. If a pressure maximum arises, large grains are more sensitive than small grains to this pressure gradient and can become trapped. Therefore, the detection of this compact emission in (sub-)mm wavelengths but not in NIR observations is indicative of dust trapping. The congregation of large grains in this dust trap can lead to the formation of planetesimals, and it additionally offers a solution to the so-called 'radial drift problem', where large ($> 1$ mm) grains rapidly drift inward toward the star \citep[e.g.][]{1977, whipple72, vdmsummary}.\\ 
In other systems with evidence for dust traps, the azimuthal asymmetries were detected at smaller radii compared to HD 34700 ($\sim 155$ AU). In Oph IRS 48, the dust trap was at a distance between 45 and 80 AU ($\pm$ 9 AU) \citep{vandermarel}. In HD 135344B, the crescent-shaped asymmetry was modeled to be located at roughly 80 AU \citep{cazzoletti}. Our cavity is significantly larger, at a distance of 155 AU, placing this among the most distant asymmetries yet detected around young stars. This larger distance to the star (or in this case, the central pair of stars) may imply different formation mechanisms. Observations also find a correlation between the mass of a star and the radius of its disk in the millimeter continuum \citep{andrews18b, tripathi17}. Simulations of disk formation show disk radii growing on a timescale as short as $10^4$ years, with larger radii around higher-mass stars \citep{bate18}. It is possible that the combined mass of the AaAb binary is sufficiently high to explain the large size of the cavity and the distance between the central binary and the compact emission region. \\
This work also reports the new detection of the dust disk around HD 34700B, the distant companion. The FWHM of the 2D Gaussian fit to the visibilities is 102 AU, which places the disk size squarely in the range established by \citet{andrews18b} for stars of this mass (disk radii between $\sim20$ and $\sim70$ AU for $M_* \approx 0.7 M_\odot$, \citet{andrews18b, monnier}). Because the disk size appears to be consistent with that of single stars similar to HD 34700B in mass and temperature, it seems that the inner AaAb binary may not have significantly impacted the evolution of the disk around HD 34700B. Additionally, the inclination of the disk around B is likely different than that of the inner circumbinary disk, implying significant misalignment (however, both inclinations are poorly constrained by this model; see footnote on Table \ref{tab:model}). Therefore, it is likely that the AaAb and B pair formed through turbulent fragmentation and/or through capture \citep{bate18} and, considering the large distance between them, evolved mostly independently of one another. No substructure is resolved in the HD 34700B disk, but future higher-resolution observations could attempt to detect substructures around the distant companion.

The most similar young star system to HD 34700 in architecture and disk morphology is HD 142527. HD 142527 is a binary composed of a $2.2 \pm 0.3 M_\odot$ star and a $0.1-0.4 M_\odot$ star in a binary with eccentricity $e = 0.5 \pm 0.2$, estimated to have an age of $5^{+8}_{-3}$ Myr \citep{verhoeff11, biller12, Lacour16}. The disk observed around the binary in HD 142527 has one of the largest eccentric circumbinary cavities known to date that is depleted in mm-size grains but not fully depleted in small dust and gas \citep{casassus12, casassus13, canovas13, Rameau12, avenhaus17}. At the edge of this cavity, there is a large crescent-shaped azimuthal asymmetry in the millimeter continuum that has been interpreted as a dust trap \citep{casassus13, casassus15, Fukagawa13}, although it could potentially be caused by an eccentric central cavity \citep{ataiee13, Price18}. The HD 142527 disk also exhibits a spiral pattern in IR observations, with spiral arms beginning at the edge of the cavity \citep{casassus12, avenhaus14, avenhaus17}. A prominent spiral arm in the IR is seen to have a radially shifted counterpart in $^{12}$CO emission, as well as other spiral arms traced by the CO gas \citep{Fukagawa06, christiaens14}. The cavity also has a double shadow, thought to be cast by a misaligned inner disk \citep{avenhaus14, marino15b}. This is similar to the darkening seen in HD 34700 AaAb's disk by \citet{Uyama20}.\\
Although HD 142527 shares many similarities with HD 34700, there are a few key differences between these systems. The semimajor axis of the HD 142527 binary is $14^{+12}_{-5}$ AU \citep{biller12}, much larger than the value of $\sim 0.25$ AU calculated for HD 34700. Additionally, the mass ratio for the central binary of HD 34700 is close to 1, whereas the mass ratio for HD 142527 is $\sim 0.1$ \citep{verhoeff11}.
The larger separation of HD 142527 translates to a larger region where stable orbits cannot occur around the central binary. Using the results of the dynamical study of binaries by \citet{Holman99} and the systems' semimajor axes, mass ratios, and eccentricities, we obtain that the smallest stable orbit around the central HD 142527 binary is $\sim 53$ AU, while the smallest stable orbit around HD 34700AaAb is $\sim 0.8$ AU. Therefore, the very different binary orbits between the two systems imply that while HD 142527's eccentric, wider separation stellar companion can explain most of the disk features, HD 34700's inner AaAb binary separation is too small to significantly affect both the disk at 100 AU and any companions with semimajor axes greater than 1 AU, potentially indicating the need  for a third body within the cavity (see Section \ref{sss:planet}.)

\subsection{Potential Origin of Observed Structure} \label{ss:causes}
\subsubsection{Central Cavity}

The inner AaAb binary is expected to truncate the disk at a radius of approximately 2 to 3 times the binary separation \citep{arty94}. The inner binary has an eccentricity of roughly 0.25 \citep{torres}, which will affect the truncation of the disk \citep{mirandalai15}. Using the relations described in \citet{mirandalai15}, we must consider the disk-binary relative inclination and the binary's semimajor axis and eccentricity to determine the expected truncation radius. The inclination of the binary is constrained spectroscopically by \citet{torres}, stating that $M_* sin^3(i) \approx 0.53 M_\odot$, which, when combined with the new masses determined by \citet{monnier} yields a central binary inclination of $i \approx 40$ or $140$ degrees. The former value is consistent with the inclination of the dust disk around the central binary ($<47^\circ$ as determined by our model in Section \ref{sec:model}, and $\sim 40 ^\circ$ from near-IR observations \citet{monnier}). We can then use the relation between cavity size and binary separation established in \citet{mirandalai15} for a given relative disk-binary inclination to estimate the cavity size that would be produced by the inner binary. Considering the inclination of $40$ or $140$ degrees, the binary semimajor axis would be $a \sim 0.25$ AU based on the spectroscopic observations of \citet{torres}. The ratio $r_{cavity} / a$ predicted by the models is 2.1 for the given inclination, mass ratio, and eccentricity, which means that the cavity that would be produced by the inner binary dynamics would be much smaller ($r_{cavity} \approx 0.53$ AU) than the observed cavity ($r \approx 175$ AU, \citet{monnier}). Therefore, even considering its eccentricity, the AaAb binary cannot be responsible for the large cavity. This motivates the introduction of a tertiary massive companion within the cavity, as discussed in Section \ref{sss:planet}.\\

\subsubsection{External Companion Dynamics}
 An external companion in a multiple system can interact with an inner circumbinary disk and potentially influence the substructures within that disk. The HD 100453 system (described in \citet{rosotti20}) has spiral arms in scattered light, as well as in the millimeter continuum, which have been determined to most likely originate from interactions with the known outer companion. In the HD 34700 system, however, the tertiary companion is much more distant than the one in HD 100453 (projected separation 1850~AU vs. 110~AU), likely too far away to truncate of the circumbinary disk and produce the spiral arms seen in scattered light \citep{rosotti20}. The CO gas emission detected around the central binary in these observations does not extend out to nearly the same distance as HD 34700B, so the only influence this distant companion could have on the inner circumbinary disk would be if HD 34700B was nearer to the central binary in the past, or if its orbit was highly eccentric. \\ 
 ``Flyby" encounters with such a perturber can cause spiral structures and asymmetries \citep{cuello20}. The perturber can also capture some of the disk material and produce tidal streams (as seen in, for example, RW Aur as described in \citet{dai15}). While past capture of circumbinary material could explain the presence of a disk around HD 34700B, there is no evidence for a tidal stream in our SMA CO observations.  However, our observed disk sizes for AaAb and B are consistent with those observed around single stars, which would argue against a close, disruptive fly-by in HD 34700's past. Nonetheless, future dynamical simulations are needed to test this scenario in detail.\\
Additionally, the fourth star HD 34700C, as well as the potential fifth member of the system suggested by \citet{monnier}, could potentially affect the dynamics of the system, but the lack of information on the true three dimensional geometry of the system limits any speculation about that. Because these companions are at distances even greater than that of HD 34700B, they are unlikely to have had significant influence on the circumbinary disk unless they have a very eccentric orbit or passed much nearer to the central binary in the distant past.

\subsubsection{Internal Companion Dynamics} \label{sss:planet}

The presence of a low mass companion within the dust cavity was proposed by \citet{monnier} to explain the presence of the cavity. They performed a targeted simulation where they found that a $50 M_{\rm Jupiter}$ companion would produce the cavity and the observed discontinuity seen in the NIR, but does not reproduce the spiral arms. To reproduce the spiral arm structure seen in NIR scattered light observations, it would require either multiple companions or an eccentric companion \citep{monnier}. An inner eccentric companion can trigger multiple outer spirals, as shown by simulations and radiative transfer modeling in \citet{calcino20}. \\
Additional observations by \citet{Uyama20} placed limits on the companion mass of $12 M_{Jupiter}$ at 0\farcs3 (or $\sim 107$ AU) and $5 M_{Jupiter}$ at 0\farcs75 (or $\sim 267$ AU). A giant companion within the limits proposed by \citet{Uyama20} could be consistent with the presence of CO and small grain emission interior to the cavity \citep{dong12, zhu12}.
The edge of a cavity produced by such a massive planet/companion would be subject to Rossby Wave Instability \citep{papapringle, lovelace}. If the disk viscosity is sufficiently low \citep{zhustone14}, this will produce a long lived anticyclonic vortex in the gas, with a gas pressure maximum at its center. This gas pressure maximum will cause dust trapping with a preference for larger grains \citep{1977, bargesommeria, birnstiel13} and potentially explain our observed major asymmetry in the millimeter continuum emission. Therefore, all observed structure in this system seems plausibly created by a massive planet or brown dwarf just interior to the cavity edge. \\
Based on calculations in \citet{Holman99}, the eccentricity of the central binary would only affect the long term stability of a companion if it is located within 0.8 AU of the central binary. This is much smaller than the cavity of interest in this system (and than the radii relevant to the mass limits in \citet{Uyama20}) so therefore it is possible for a companion to be at a stable orbit at a wide range of semimajor axes within the cavity. Detailed dynamical simulations are needed to test whether a planet within the mass limits imposed by \citet{Uyama20} would be capable of reproducing all of the observed features, including the cavity size, CO and small dust inward filtering, and high contrast azimuthal mm-dust trapping.\\
If a planet is invoked as the cause of the cavity and asymmetry, then it indicates that the formation of a massive companion as distant as 160 AU must have taken place rapidly within the system age of 5 Myr, unless the orbit of the companion is highly eccentric. This rsigniapid companion formation could have been facilitated by a very massive protostellar disk, as is commensurate with the binary pair of massive stars at its center.\\

 both\subsubsection{Gravitational Instability}
If the protostellar disk indeed was very massive in the past, it may have been gravitationally unstable (see \citet{kratterlodato} for a review of gravitational instability). \\
The spiral arms observed in scattered light by \citet{monnier} are a specific prediction of gravitational instability models \citep{dipierro14, hall19}. However, 
\citet{monnier} argued that, based on radiative transfer modeling, that the Toomre Q parameter is greater than 25 everywhere in the disk and therefore that it is unlikely the disk is gravitationally unstable at present. 
\citet{kratterlodato} states that disks are generally not gravitationally unstable unless the $\frac{M_d}{M_*}$ ratio is greater than $10^{-2}$.
To calculate the total disk mass, we use our calculated CO mass limits from Section \ref{ss:comass}, assume a  $H_2 / CO$ ratio of $10^4$, and add the dust mass by assuming a gas-dust ratio (by mass) of 10. The lower limit of the total estimated disk mass is $3.6 \times 10^{-5} M_\odot$, and the upper limit is $4.3 \times 10^{-4} M_\odot$. This yields, for this system, $\frac{M_d}{M_*} \geq 10^{-4}$ using the lower limit on the CO mass, or $\frac{M_d}{M_*} \geq 10^{-3}$ using the upper limit on the CO mass. This strongly implies that the disk is not gravitationally unstable at present, unless it is more massive than our assumptions allow. Even if the disk is not gravitationally unstable at present, it may have been gravitationally unstable in the past when it was more massive. This could cause the rapid and early formation of the massive companion proposed by \citet{monnier} to explain the observed substructure. 

\section{Summary and Conclusions} \label{sec: conc}
In this paper, we present new high-resolution ($\sim 0\farcs5$) SMA observations of the HD 34700 multiple system. We discover azimuthally asymmetric  emission in the 1.3mm continuum around the central AaAb binary, which we interpret as evidence for dust trapping. We also report the discovery of dust emission associated with a protoplanetary disk around the distant companion HD 34700B. We determine the inclination of the AaAb circumbinary disk is consistent with being aligned with the inclination of the plane of the central binary as spectroscopically constrained by \citet{torres}. \\
The substructures observed both in (sub-)mm and NIR observations \citep{monnier} can be explained by the presence of a companion in the circumbinary disk (or companions) that must have formed within 5 Myr. To form a massive companion so rapidly, it is possible that the disk around HD 34700AaAb was massive enough to be gravitationally unstable in the past. Our calculated disk mass, inferred from CO emission, implies that the disk is not gravitationally unstable at present. As mentioned in \citet{monnier}, a deeper search for point sources within $\sim 170$ AU from the central binary would be necessary to confirm the presence of a young companion that could potentially explain the dust trap observed in millimeter wavelengths and the spiral arms seen in scattered light. The observed structures indicate that this system is an interesting candidate for future high-sensitivity high-angular resolution ALMA observations to enable a better characterization of the disk and better constraints on the origin of the observed disk features.  \\
\\
\textbf{Acknowledgments}

J.D.M./E.A.R. acknowledge NSF AST 1830728. This work has made use of data from the European Space Agency (ESA) mission Gaia \citep{gaia}, processed by the Gaia Data Processing and Analysis Consortium (DPAC). Funding for the DPAC has been provided by national institutions, in particular the institutions participating in the Gaia Multilateral Agreement. This research has also made use of the SIMBAD database \citep{wenger2000}, operated at CDS, Strasbourg, France, and NASA's Astrophysics Data System Bibliographic Services.\\
The authors also wish to acknowledge the cultural significance of the summit of Mauna Kea to the indigenous Hawaiian people. We are fortunate to have the privilege to observe from this mountain. We also thank the anonymous referee for their useful commentary.

\singlespace
\bibliographystyle{apj}
\bibliography{diskbib}

\end{document}